\documentclass[aps,showpacs,preprint,superscriptaddress]{revtex4}
\usepackage{graphicx}
\usepackage{subfigure}
\usepackage{float}
\usepackage{bm}
\usepackage{txfonts}
\usepackage{array}

\begin{document}

\title{Modulation effect in multiphoton pair production}
\author{Ibrahim Sitiwaldi}
\affiliation{Key Laboratory of Beam Technology and Materials Modification of the Ministry of Education, and College of Nuclear Science and Technology, Beijing Normal University, Beijing 100875, China}
\author{Bai-Song Xie \footnote{Corresponding author. Email address: bsxie@bnu.edu.cn}}
\affiliation{Key Laboratory of Beam Technology and Materials Modification of the Ministry of Education, and College of Nuclear Science and Technology, Beijing Normal University, Beijing 100875, China}
\affiliation{Beijing Radiation Center, Beijing 100875, China}

\date{\today}
\begin{abstract}
We investigate the electron-positron pair production process in an oscillating field with modulated amplitude in quantum kinetic formalism. By comparing the number density in field with and without modulation, we find that the pair production rate can be enhanced by several orders when the photon energy just reach the threshold with the help of shifted frequency due to modulation. We also detect the same effect in a pulse train with subcycle structure. We demonstrate that the frequency threshold can be lowered by frequency of pulse-train due to modulation effect. We also find that the momentum distribution for $N$-pulse train can reach $N^2$ times the single pulse at the maximum value and the number density as a function of pulse number follows the power laws with index $1.6$ when the modulation effect is maximized.

\textbf{key words}: pair production, modulation effect, pulse train
\end{abstract}
\maketitle

\section{Introduction}

The electron-positron pair production from vacuum under the strong external field is one of remarkable predictions of the quantum electrodynamics (QED). After pioneering works of Sauter \cite{Sauter}, Heisenberg and Euler \cite{Heisenberg}, and Schwinger \cite{Schwinger} a large number of investigations are dedicated to study on vacuum pair production through employing different methods such as proper time technique \cite{propertime1, propertime2, propertime3}, Wentzel-Kramers-Brillouin (WKB) approximation \cite{WKB}, worldline instanton technique \cite{world1, world2}, quantum field theoretical simulation \cite{DiracRecent,CreationDynamics} as well as the quantum kinetic method \cite{kinetic1, kinetic2,kinetic3,kinetic4}. The experimental verification of vacuum pair production is still remaining unavailable so far due to Schwinger threshold of external electric field $E_{cr}=m^2c^3/e\hbar=1.32\times10^{18}\mathrm{V/m}$ ($m$ and $-e$ denote mass and charge of electron, respectively.), which is too high to achieve in the laboratory at present.

With the rapid development of laser technology in recent years, it is being expectable that the field strength of experimental facility may be more approaching to the Schwinger threshold in the near future, for example, the European extreme-light-infrastructure (ELI) program is now advancing \cite{ELI}. Motivated by this, various schemes have been proposed to support the upcoming possible experiments in recent years. A strongly enhanced pair production rate was presented in dynamically assisted Schwinger mechanism where a rapid oscillating electric field is superimposed onto a slowly varying one \cite{dynamical1,dynamical2,dynamical3}. The study on time-domain multi-slit interference effect in alternating sign $N$-pulse electric field \cite{slit} showed that the maximum of central longitudinal momentum can reach $N^2$ times the single pulse value. The pair production in a short pulse with subcycle structure was studied  \cite{kinetic0} to present momentum spectrum extremely sensitive to subcycle dynamics. The accurate study on pair production in a pulsed electric field with subcycle structure \cite{emass} detected the signature of effective mass of electrons and positrons in the given strong electric field, where the effective mass is a little higher than real mass so that the frequency threshold rises a little correspondingly. These findings and schemes advance greatly the topic of pair production in strong field, more details can be seen in \cite{nuriman, dynamical0, stimulated, lzl, bif1,bif2,bif3}.

In present letter we introduce a scheme where the amplitude of a high frequency oscillating electric field is modulated. Amplitude modulating change the dynamics of oscillating field in the following two aspects. On the one hand the frequency of oscillating field is shifted up by modulation frequency, which may have a positive influence on the production rate, on the other hand the field strength is decreased to suppress the production rate in multiphoton regime. Therefore it is desirable to show overall influence of the amplitude modulation on pair production process. In this paper, we study the production rate and momentum distribution for different parameters in a high frequency oscillating field with amplitude modulated by a sinusoidal signal. The study is to highlight how the modulation effect influences the pair production. Furthermore, we investigate how the modulation effect can lower the frequency threshold in a more realistic field configuration-pulse train with subcycle structure. We employ quantum Vlasov equation in the kinetic formalism and all quantities are working on the natural units ($\hbar=c=1$) in this paper.

The paper is organized as follows. In Sec.\ref{method} we introduce the quantum Vlasov equation for a completeness. In Sec.\ref{result} we get the numerical results and necessary theoretical analysis. In the last section we provide a brief conclusion.

\section{Theoretical formalism based on quantum Vlasov equation}\label{method}

The spectral information of created particles in an external field is encoded in the distribution function $f(\mathbf{p}, t)$. The equation of motion for $f(\mathbf{p}, t)$ can be derived from canonical quantization with fully quantized spinor and the electromagnetic field as a classical background \cite{kinetic3}.  We are only interested in subcritical field strength regime $E\ll E_{cr}$ where created particle density is so low that the collision effect and self consistent field current due to created particles can be neglected. Since the achievable spatial focusing scale is orders of magnitude larger than the Compton wavelength of electrons we ignore any spatial dependence, and we also ignore the magnetic field. With these simplifications, the quantum Vlasov equation for $f(\mathbf{p}, t)$ reads:
\begin{equation}
\dot{f}(\mathbf{p}, t) =
\frac{1}{2}q(\mathbf{p}, t)\int_{-\infty}^{t}dt' q(\mathbf{p}, t')
[1-2f(\mathbf{p}, t')]\cos[2\Theta(\mathbf{p}, t', t)],  \label{vlasov}
\end{equation}
where $f(\mathbf{p}, t)$ accounts for both spin directions due to absence of magnetic fields. Here, $q(\mathbf{p}, t)=eE(t)\varepsilon_{\bot}/\omega^{2}(\mathbf{p}, t)$ and $\Theta(\mathbf{p}, t', t)=\int_{t'}^{t}\omega(\mathbf{p}, \tau)d\tau$ with quantities as the electron/positron momentum $\mathbf{p}=(\mathbf{p}_{\bot}, p_{\parallel})$,  transverse energy squared $\varepsilon_{\bot}^{2}=m_{e}^{2}+p_{\bot}^{2}$, the total energy squared
$\omega^{2}(\mathbf{p}, t)=\varepsilon_{\bot}^{2}+p_{\parallel}^{2}$, and the longitudinal momentum $p_{\parallel} = P_{3}-eA(t)$.
This equation may be expressed as a linear, first order, ordinary differential equation system (ODEs) \cite{kinetic4} for the convenience of numerical treatment:
\begin{eqnarray}
\dot{f}(\mathbf{p}, t) & = & \frac{1}{2}q(\mathbf{p}, t)g(\mathbf{p}, t), \\
\dot{g}(\mathbf{p}, t) & = & q(\mathbf{p}, t)[1-2f(\mathbf{p}, t)]
-2\omega(\mathbf{p}, t)w(\mathbf{p}, t),  \\
\dot{w}(\mathbf{p}, t) & = & 2\omega(\mathbf{p}, t)g(\mathbf{p}, t).
\label{ode}
\end{eqnarray}
The term $g(\mathbf{p},t)$, i.e. the integral part of Eq.(\ref{vlasov}) constitutes an important contribution to the source of pair production, where the quantum statistics character is represented by the term $[1-2f(\mathbf{p},t)]$ due to the Pauli exclusive principle. The term $w(\mathbf{p},t)$ denotes a countering term to pair production, which is associated to the pair annihilation in pair creation process to some extent. The last one of ODEs means that the more pairs are created, the more pairs are annihilated probably in pair creation process. Note that the studied system has a typical non-Markovian character.

By numerically solving this ODEs with the initial conditions $f(\mathbf{p}, -\infty)=g(\mathbf{p}, -\infty)=w(\mathbf{p}, -\infty)=0$, we can obtain spectral information $f(\mathbf{p}, t)$ of the created particles for any given spatially homogenous, time-dependent electric field. The number density $n(t)$ of created particles can be obtain easily from $f(\mathbf{p}, t)$:
\begin{equation}
n(\infty)=2\int\frac{d^{3}\mathbf{p}}{(2\pi)^{3}}f(\mathbf{p}, \infty),
\label{number}
\end{equation}
where the factor $2$ comes from the degeneracy of electrons. It is need to remind that the particle interpretation of $f(\mathbf{p}, t)$ is invalid in the presence of external field, it can be considered as the distribution function for real particles only at asymptotic times $t=\pm\infty$.

\section{Numerical Results}\label{result}

In order to reveal the physical mechanism briefly as well as for the simplicity of numerical calculations, we just consider the problem in the one-dimensional momentum space for created electron-positron pair.

\subsection{Pair production in a high frequency oscillating field with amplitude modulated by a sinusoidal signal}

\begin{figure}[H]\suppressfloats
\includegraphics [width=8.5cm]{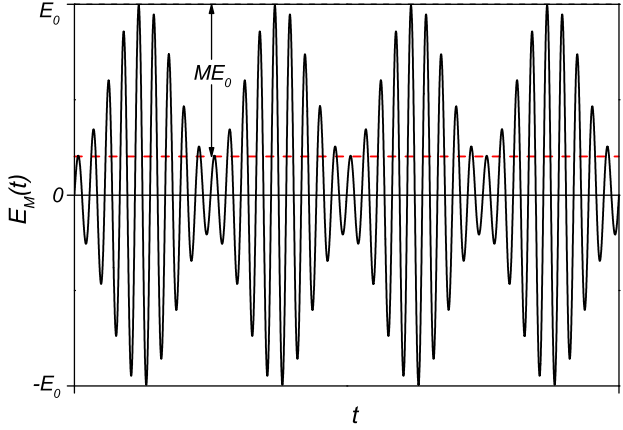}
\caption{\label{modulation1} High frequency field with amplitude modulated by a sinusoidal signal with modulation degree $M$.}
 \end{figure}

The spatially homogenous, time-dependent electric field in a given direction can be expressed as $\mathbf{E}(t)=-\dot{\mathbf{A}}(t)=(0,0,E(t))$. The transverse momentum of created particles which is perpendicular to electric field is fixed as $p_{\perp}=0$ as mentioned above. We introduce a high frequency oscillating field (carrier) with frequency $\omega_c$ and amplitude $E_0$ modulated by a sinusoidal signal with modulation frequency $\omega_m$ ($\omega_m\ll\omega_c$),
\begin{equation}\label{am1}
E_M(t)=(1-M\frac{1+\cos(\omega_mt)}{2})E_0\sin(\omega_ct),
\end{equation}
where $0\leq M\leq1$ denotes modulation degree ($M=0$ for no modulation and $M=1$ for full modulation) as displayed in Fig.\ref{modulation1}.

Now Eq.(\ref{am1}) can be expanded as $E_M(t)=(1-\frac{M}{2})E_0\sin(\omega_ct)-\frac{M}{4}E_0\sin(\omega_+t)-\frac{M}{4}E_0\sin(\omega_-t)$, where $\omega_{\pm}=\omega_c\pm\omega_m$. It indicates that modulating amplitude results in frequency shift, from which one can expect a possible positive influence on the pair production rate. On the other hand the average  strength (power) of electric field is suppressed by a factor of $\int_0^{2\pi/\omega_m}E^2_M(t)dt/\int_0^{2\pi/\omega_m}E^2_0(t)dt=\frac{3}{8}(\frac{4}{3}-M)^2+\frac{1}{3}$  due to modulation, which may have a negative influence on the pair production rate.

\begin{figure}[H]\suppressfloats
\includegraphics [width=8.5cm]{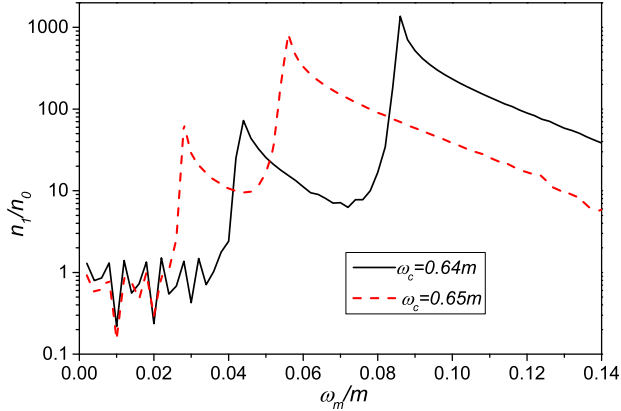}
\caption{\label{omegam} (color online). The increasing factor $n_1/n_0$  as a function of modulation frequency $\omega_m$ for two value of carrier frequency $\omega_c=0.64m$ (black solid line) and $\omega_c=0.65m$ (red dashed line).}
 \end{figure}

To investigate the overall influence of amplitude modulation on the pair production rate, we compare produced number density with and without modulation for full frequency space. We find that the pair production rate can be enhanced due to modulation for carrier frequency near the frequency threshold of multiphoton process.  We discuss the results corresponding to three-photon regime here, without losing generality. In Fig.\ref{omegam} we display the increasing factor $n_1/n_0$  as a function of modulation frequency $\omega_m$ for carrier frequencies  $\omega_c=0.64m$ and $\omega_c=0.65m$ with field strength $E_0=0.1E_{cr}$, where $n_1$ and $n_0$ denote the produced numbers density for full modulation $M=1$ and no modulation $M=0$, respectively. To avoid trivial effects we switch electric field on and off exponentially within time $t_{switch}=100\pi\tau_0$ ($\tau_0=1/m$ denotes the electron Compton time), and chose the duration time long enough $t_d=1000\pi\tau_0$ where total calculation time is $2t_{switch}+t_d$. The modulation frequency is scanned with a step of $0.002m$.

It can be seen that the increasing factor $n_1/n_0$ strongly depends on modulation frequency. There are two main peaks for each case where the production rate is enhanced by several orders due to modulation. The first peaks for $\omega_c=0.64m$ and $\omega_c=0.65m$ are at $\omega_m=0.044m$ and  $\omega_m=0.028m$ with increasing factor $72$ and $65$, respectively. Both of this peaks satisfy a relation as $\omega_c+2\omega_+=2.006m$. The second peaks for $\omega_c=0.64m$ and $\omega_c=0.65m$ are at $\omega_m=0.088m$ and $\omega_m=0.056m$ with increasing factor $1364$ and $445$, respectively. These peaks also satisfy a relation as $2\omega_c+\omega_+=2\omega_++\omega_-=2.006m$. This result indicates that the pair production rate can be enhanced due to modulation when the gap between carrier frequency and threshold is supplemented by modulation frequency. The threshold value here is $2.006m$ instead of $2m$ can be explained by effective mass $m_*$ \cite{emass}, where $2m_*=2.006m$ agree with the result of analytic calculation: $m_*=m\sqrt{1-\langle A^\mu A_\mu\rangle e^2/m^2} \approx 1.0023m$.

\begin{figure}[H]\suppressfloats
\includegraphics [width=8.5cm]{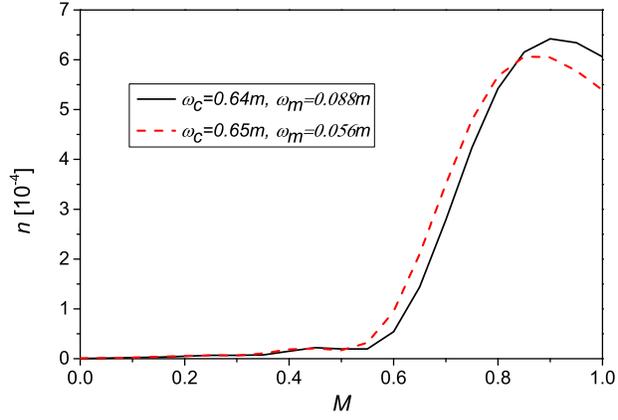}
\caption{\label{modul} Produced number density as a function of modulation degree $M$ for  two cases of carrier and modulation frequencies $\omega_c=0.64m$, $\omega_m=0.088m$ (black solid line) and $\omega_c=0.65m$, $\omega_m=0.056m$ (red dashed line). Other parameters are the same with in Fig.\ref{omegam}.}
 \end{figure}

In Fig.\ref{modul} we display produced number density as a function of the modulation degree for two cases corresponding to the main peaks in Fig.\ref{omegam}. There are no clear enhancement before modulation degree $M$ reachs $0.5$ due to the fact that the amplitude of upshifted frequency component $E_{cr}M/4$ is not strong enough to participate the multiphoton process. The produced number density dramatically increase when $M$ exceeds $0.6$. It indicates that the shifted frequency component begin to play a role. The production rate maximized at $M=0.9$ and $M=0.85$ are seen for two cases, respectively. This non-monotonic behaviour can be explained by the competition between decreasing total field strength and increasing component of upshifted frequency.

\begin{figure}[H]\suppressfloats
\includegraphics [width=17cm]{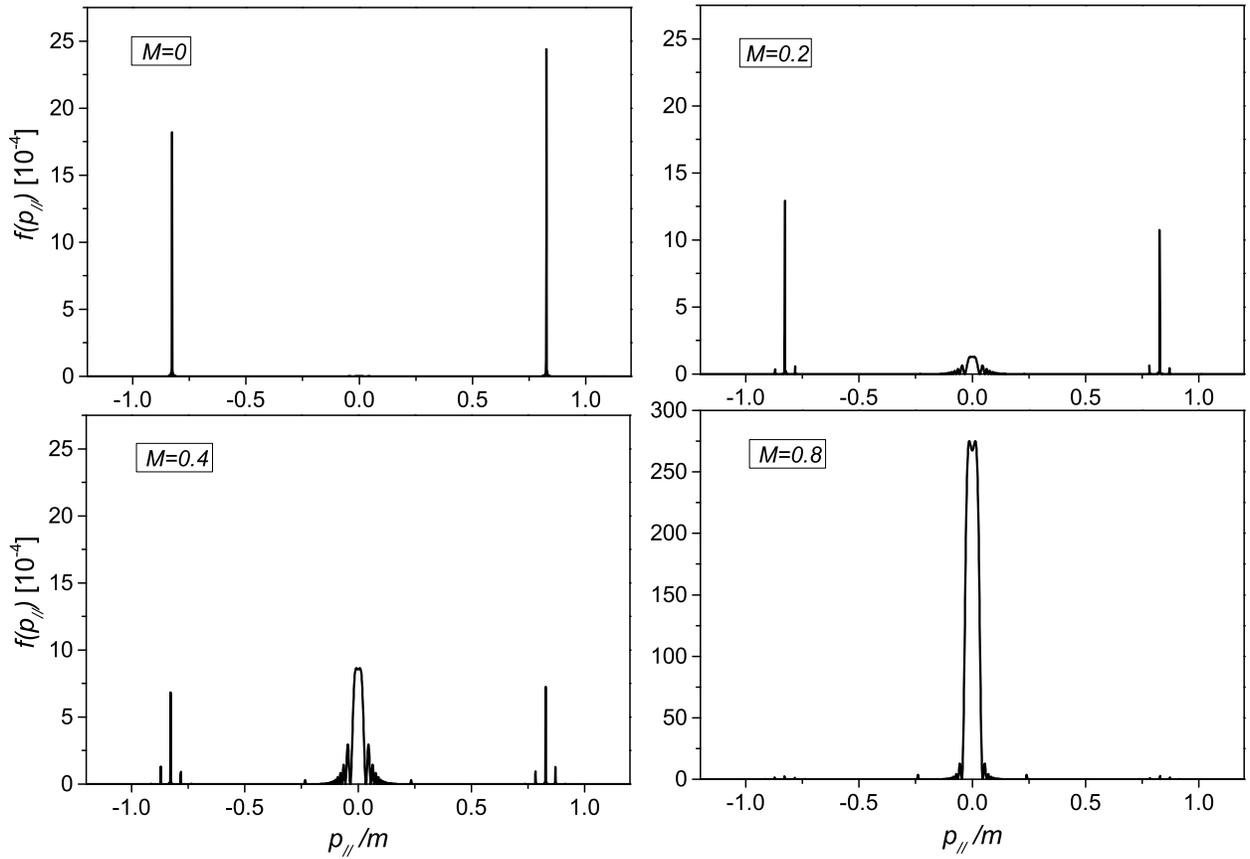}
\caption{\label{momentum} The longitudinal momentum distribution for different modulation degrees.}
\end{figure}

The longitudinal momentum distributions for different modulation degrees are displayed in Fig.\ref{momentum}, where carrier frequency $\omega_c=0.65m$ and modulation frequency $\omega_m=0.056m$, other parameters are the same as in Fig.\ref{omegam}. For $M=0$, i.e. the case without modulation, the momentum distribution is peaked at $p_{\parallel}=\pm0.83m$, satisfying the energy-momentum relation for four-photon process, $4\omega_c=2\sqrt{p_{\parallel}^2+m^2}$. As $M$ increases, for example, when $M=0.2$ and $M=0.4$, the peaks at  $p_{\parallel}=\pm0.83m$ become lower and some new peaks appear around of $p_{\parallel}=0$, which satisfy the energy-momentum relation in threshold value ($p_{\parallel}\approx0$) with effective mass for three-photon processes $2\omega_c+\omega_+=2\omega_+ + \omega_-=2m_*$. The small peaks at $p_{\parallel}=\pm0.83m$ are still corresponding to the energy-momentum relation for four-photon processes but with $4\omega_-=2\sqrt{p_{\parallel}^2+m^2}$ and $4\omega_+=2\sqrt{p_\parallel^2+m^2}$, respectively. As $M$ increases further almost only one peak around $p_{\parallel}=0$ is remained, for example, see the case of $M=0.8$, which indicates that the hole process is dominated completely by the three-photon processes due to modulation. This result also agrees with the fact in Fig.\ref{modul} that the number density reach the maximum after $M$ exceeds $0.8$.

\subsection{The modulation effect among pulses in a pulse train}

Now we consider a $N$-pulse train with subcycle structure where the field is a superposition of a serial pulses as
$E(t)=\sum_{n=1}^{N}E_n(t)$ as displayed in Fig.\ref{pulse} when $N=3$. Here the $n^{th}$ pulse $E_n(t)$ can be represented as:
\begin{equation}
E_n(t)=E_0e^{-\frac{(t-nT_m)^2}{\tau^2}}\sin(\omega_ct),
\label{pulse}
\end{equation}
where $E_0$ denotes field strength of each pulse, $\omega_c$ denotes the carrier frequency, $\tau$ denotes the pulse duration and $T_m$ denotes the time delay of pulse train and also it is associated to the peak positions of $n^{th}$ pulse field as $nT_m$. Note that we fix the $\tau$ for each pulse so that the modulation by Gaussian envelop for each pulse are the same but the different delay time $T_m$ now plays a role of modulation of pulse train. Thus we can define a pulse-train frequency as $\omega_m=2\pi/T_m$ as the modulation frequency of pulse-train when $N>1$. Obviously there is no modulation of pulse-train for a single pulse when $N=1$.

\begin{figure}[H]\suppressfloats
\includegraphics [width=8.5cm]{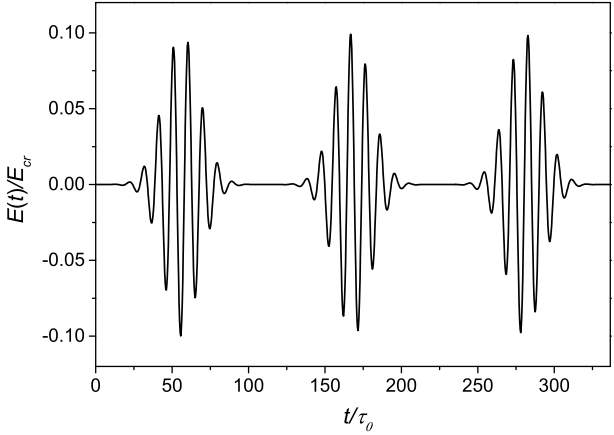}
\caption{\label{pulse} The illustration of pulse train with subcycle structure when $N=3$ for $\omega_c=0.65m$, $\omega_m=0.056m$, $\tau=16\tau_0$ and $E_0=0.1E_{cr}$.}
 \end{figure}

\begin{figure}[H]\suppressfloats
\includegraphics [width=8.5cm]{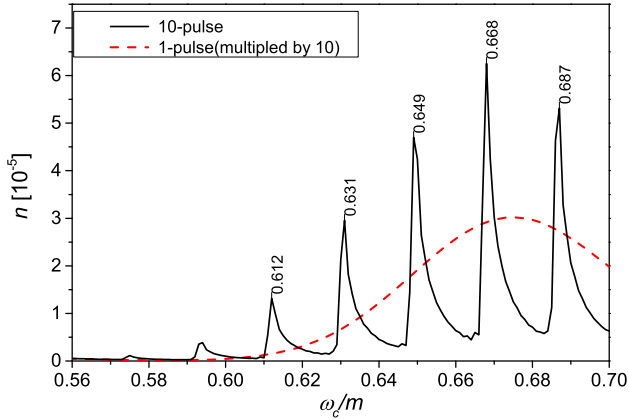}
\caption{\label{omegac} (color online). The produced number density as a function of carrier frequency in a pulse train with subcycle structure for one pulse when $N=1$ (dashed line) and ten pulses when $N=10$ (solid line), respectively.}
 \end{figure}

We display the produced number density as a function of carrier frequency in a pulse train with subcycle structure for one pulse ($N=1$) and ten pulses ($N=10$) in Fig.\ref{omegac} with parameters $\omega_m=0.056m$, i.e. $T_m=112.2\tau_0$, $E_0=0.1E_{cr}$, $\tau=16\tau_0$.

The curve for one pulse (amplified by 10 times for convenience) is peaked at the ordinary frequency threshold $\omega_c=0.67m$ for the three-photon process as one may expect. However, the curve for ten pulse is peaked at several places with the same distance. We suppose this field can also be composed of three single oscillating fields with frequencies $\omega_{\pm}=\omega_c\pm\omega_m$ and $\omega_c$ analogous to the case in the last subsection. The peak at $\omega_c=0.668m$ is corresponding to the ordinary frequency threshold of three-photon process, with the relation $3\omega_c=2m_*$ (we take the effective mass $m_*=1.002m$ in this subsection), and the peak at $\omega_c=0.649m$ is corresponding to $2\omega_c+\omega_+=2\omega_++\omega_-=2m_*$, where the carrier frequency is lower than threshold frequency by $0.019m$ while the number density is about still of $80\%$ of that in case of threshold frequency. These peaks at $\omega_c=0.631m$, $\omega_c=0.612m$ and $\omega_c=0.687m$ are also corresponding to $\omega_c+2\omega_+=2m_*$, $3\omega_+=2m_*$ and $2\omega_c+\omega_-=2\omega_-+\omega_+=2m_*$, respectively. The difference in peak heights can be explained by different strength of each frequency component as we expanded in last subsection. This result indicates that the frequency threshold can be lowered by a shifted frequency due to modulation among pulses in a pulse train.

\begin{figure}[H]\suppressfloats
\includegraphics [width=8.5cm]{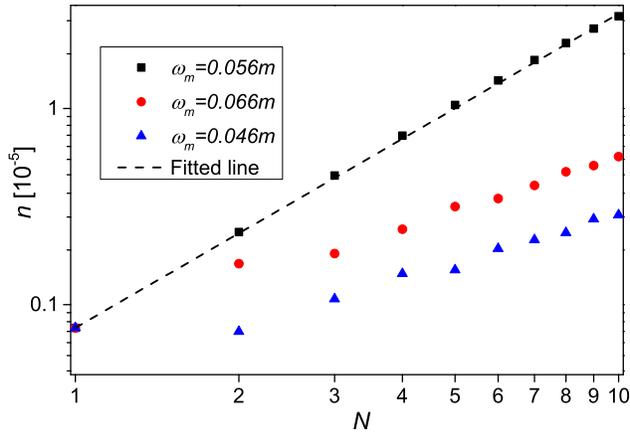}
\caption{\label{npulse} (color online). The number density in a pulse train with subcycle structure as a function of pulse number $N$ for $\omega_c=0.65m$  with different modulation frequencies $\omega_m$, the fitted line is $n_N=n_1N^{1.6}$. }
\end{figure}

The produced number density as a function of pulse number $N$ for $\omega_c=0.631m$ and $\omega_m=0.056m$ is displayed in  Fig.\ref{npulse}. As a comparison, the results for alternative modulation frequencies of $\omega_m=0.046m$ and $\omega_m=0.066m$ are shown also. Other parameters are the same as in Fig.\ref{omegac}.

The number densities for three values of $\omega_m$ are the same when $N=1$. This is not surprising since that for a single pulse the modulation among pulses is absent and the different $T_m$ just shift the position of peak field. However, as the pulse number increases, the number density is greatly increased for three sets of $\omega_m$ but the increasing for $\omega_m=0.056m$ is most remarkable because the $\omega_c+2\omega_+=2m_*$ holds in this case in comparisons with either $\omega_m=0.046m$ or $\omega_m=0.066m$. Thus it represents a typical significance of modulation effect for the pair production in multiphoton regime. Especially the number density as a function of pulse number follows the power laws with index $1.6$ for $\omega_m=0.056$ as represented with the dashed line, which is very similar to the result in the case of interference effect with dynamically assisted Schwinger mechanism \cite{lzl}. In the other word the power law presented in publication before and here can be explained briefly as a result of multiphoton matching condition of effective frequency for pair production. Obviously this effective frequency is constituted by the original field carrier frequency and the modulation effect among the pulses which is associated to the delay time of pulse train.

\begin{figure}[H]\suppressfloats
\includegraphics [width=8.5cm]{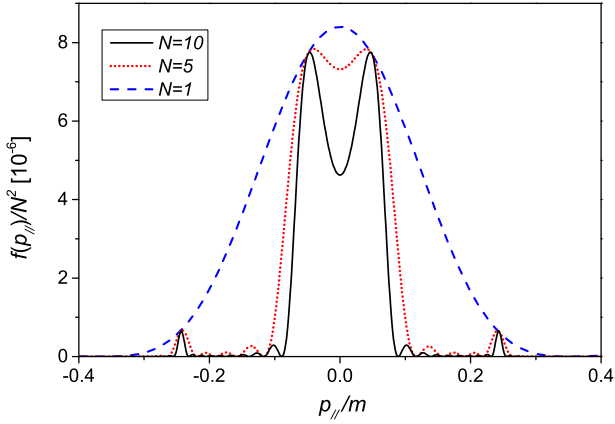}
\caption{\label{mom} (color online). The longitudinal momentum distributions normalized by square of pulse number for different pulse numbers of $N=1$ (blue dashed line), $N=5$ (red dotted line) and $N=10$ (black solid line).}
\end{figure}

The longitudinal momentum distributions normalized by square of pulse number for the case of $\omega_c=0.631m$ and $\omega_m=0.056m$ for different pulse numbers are displayed in Fig.\ref{mom}, other parameters are the same as in Fig.\ref{omegac}. Three curves are identical at the main peaks located at $p_\parallel=\pm0.05m$ and $p_\parallel=\pm0.24m$, indicating that the momentum distribution in maximum value for $N$-pulse can reach $N^2$ times the single pulse value. This result is similar to the case of time-domain multi-slit interference effect \cite{slit}. The small oscillations between the main peaks for the cases of $N=5$ and $N=10$ can be regard as a result of interference between pulses, but not very strong as in \cite{slit}.

\section{Conclusion}

The electron-positron pair production process in the high frequency oscillating field with amplitude modulation is investigated, where both of the problems that the field amplitude is modulated by a simple sinusoidal signal and field structure is modulated by a time-delay pulse train are considered, respectively.

Our main findings and conclusions are the follows:

$\bullet$ The number density with and without modulation is compared for different carrier and modulation frequencies for the case of sinusoidal modulation. It is demonstrated that the pair production rate can be significantly enhanced due to modulation when the gap between threshold frequency and carrier frequency is just supplemented by the modulation frequency. The number density depends non-monotonically on the modulation degree. The significant change of the order of multiphoton process due to modulation is represented by the momentum distribution.

$\bullet$ The modulation effect in a pulse train with subcycle structure is also investigated. It is found that the frequency threshold can be lowered due to the modulation effect among pulses so that the carrier frequency need not just reach the ordinary threshold $\omega_{cr}=2m_*/n$ to trigger $n^{th}$ order multiphoton process. It was also demonstrated that number density as a function of pulse number is fitted by the power laws with index $1.6$ and the momentum distribution in maximum for $N$-pulse can reach $N^2$ times the single pulse value when the modulation effect is maximal, i.e. that is matching the multiphoton condition of pair production.

The present study indicates that there is an effective frequency which is higher than laser frequency in a field configuration where the laser strength is periodically changing. It is reasonable to expect further investigations on the same effect in other scheme such as combined pulse or bi-frequent pulse train in future work. Certainly the modulation effect on pair production in full momentum space of created pairs is also worthy to study.

\begin{acknowledgments}
We enjoyed several helpful discussions with Moran Jia and Feng Wan. This work was supported by the National Natural Science Foundation of China (NSFC) under Grant No. 11475026.
\end{acknowledgments}

\end{document}